\documentclass[useAMS,usenatbib]{mn2e}
\usepackage{placeins}
\usepackage{epsf}
\usepackage{amssymb}
\usepackage{amsmath}
\usepackage[usenames]{color}
\usepackage{graphicx}

\def\deg{$^{\circ}$}

\newcommand{\be}{\begin{equation}}
\newcommand{\ee}{\end{equation}}

\def\***#1{\textbf{\textsf{*** #1 ***}}}

\title[110 years ago]{Not that long time ago in the nearest galaxy: 3D slice of  molecular gas revealed by a 110 years old flare of Sgr A$^*$.}
\author[Churazov et al.]{E.~Churazov,$^{1,2}$ 
  I.~Khabibullin,$^{1,2}$  R.~Sunyaev,$^{1,2}$  G.~Ponti$^3$
 \newauthor \\
$^1$ Max-Planck-Institut f\"ur Astrophysik, Karl-Schwarzschild-Strasse 1, 85741
Garching, Germany\\
$^2$ Space Research Institute (IKI), Profsoyuznaya 84/32, Moscow 117997, 
Russia\\
$^3$ Max-Planck-Institut f\"ur extraterrestrische Physik, Giessenbachstrasse 1, Garching, 85748, Germany \\
}

\begin{document}
%\date{Accepted ????????????;  Received ????????????; in original form ????????????}

\pagerange{\pageref{firstpage}--\pageref{lastpage}}
%\pubyear{2009}

\maketitle

\label{firstpage}
\begin{abstract}
A powerful outburst of X-ray radiation from the supermassive black
hole Sgr~A* at the center of the Milky Way is believed to be
responsible for the illumination of molecular clouds in the central
$\sim$100 pc of the Galaxy
\citep[][]{1993ApJ...407..606S,1996PASJ...48..249K}. The
reflected/reprocessed radiation comes to us with a delay corresponding
to the light propagation time that depends on the 3D position of
molecular clouds with respect to Sgr~A*. We suggest a novel way of determining the age of the outburst and positions of the clouds by studying characteristic imprints left by the outburst in the spatial and time variations of the
reflected emission. We estimated the age of the outburst that illuminates the Sgr~A molecular complex to be
$\sim 110$ yr. This estimate implies that we see the gas located
$\sim$10 pc further away from us than Sgr~A*. If the Sgr~B2 complex is
also illuminated by the same outburst, then it is located $\sim$130 pc
closer than our Galactic Center.  The outburst was short (less than
a few years) and the total amount of emitted energy in X-rays is
$\displaystyle \sim 10^{48}\rho_3^{-1}$~erg, where $\rho_3$ is the
mean hydrogen density of the cloud complex in units of $10^3~{\rm
  cm^{-3}}$. Energetically, such fluence can be provided by a partial
tidal disruption event or even by a capture of a planet. Further
progress in more accurate positioning and timing of the outburst
should be possible with future X-ray polarimetric observations and
long-term systematic observations with Chandra and XMM-Newton. A few hundred-years long X-ray  observations would provide a detailed 3D map of the gas density distribution in the central $\sim 100$~pc region.
\end{abstract}


%
%________________________________________________________________

\sloppypar

\section{Introduction}
While the luminosity of the supermassive black hole in the Milky Way
is currently many orders of magnitudes lower than its Eddington limit
\citep[see, e.g.,][for review]{2010RvMP...82.3121G}, it is plausible
that in the recent past the black hole was much more luminous \citep[see][for review]{2013ASSP...34..331P}. In
particular, a prominent spectral component, which is reminiscent of a spectrum
formed when X-rays are ``reflected'' by cold gas, has been found from
several molecular clouds within the central $\sim$1\deg region around the
Galactic center
\citep[e.g.,][]{1993ApJ...407..606S,1996PASJ...48..249K,2004A&A...425L..49R}. Furthermore,
the intensity of this reflected emission varies significantly on time
scales of several years
\citep[e.g.,][]{2007ApJ...656L..69M,2008PASJ...60S.201K,2009PASJ...61S.241I,2010ApJ...714..732P,2010ApJ...719..143T,2013A&A...558A..32C}, excluding
the possibility that the observed emission is induced by a steady
population of low energy cosmic rays, electrons or protons \citep[see,
  e.g.,][]{2002ApJ...568L.121Y,2012A&A...546A..88T} interacting with molecular gas. A
population of numerous persistent X-ray sources can also be excluded,
since a large equivalent width of the neutral iron fluorescent line
suggests that the primary emission is not contributing to the observed
spectra.  At present, the most plausible scenario assumes that an outburst (or several
outbursts) of the Sgr A$^*$ emission few hundred years ago is
responsible for the observed emission, although
one cannot exclude another very bright transient source (or, several
such sources) as a culprit for individual clouds \citep[see, e.g.,][]{1991ApJ...383L..49S,1993ApJ...407..752C,1996ApJ...464L..71C,1998MNRAS.297.1279S,1999hxra.conf..102S}. A number of observational and theoretical
studies discuss possible ways of proving that primary radiation is
indeed coming from Sgr A$^*$ and implications for the black hole past
outbursts \citep[see, e.g., recent studies][and references therein]{2015ApJ...814...94M,2015ApJ...815..132Z,2016A&A...589A..88M}. 

Here, we compare spatial and temporal variations of the ``reflected''
flux to infer the time of the outburst and 3D positions of illuminated
clouds. In \S\ref{sec:maps}, we describe the construction of 
reflected emission maps. We then use these maps to calculate the structure
functions in space and time domains (\S\ref{sec:sf}), and infer the
time of the outburst and 3D positions of illuminated clouds. The
implications of our findings are discussed in \S\ref{sec:dis}. The
final section summarizes our results.

Throughout the paper we assume the distance to Sgr~A$^*$ of 8~kpc, $1'$ corresponds to 2.37~pc.

\section{Reflection component maps}
\label{sec:maps}
\begin{figure*}
    \includegraphics[trim= 0mm 0cm 0mm 0cm, width=1\textwidth,clip=t,angle=0.,scale=
  0.95]{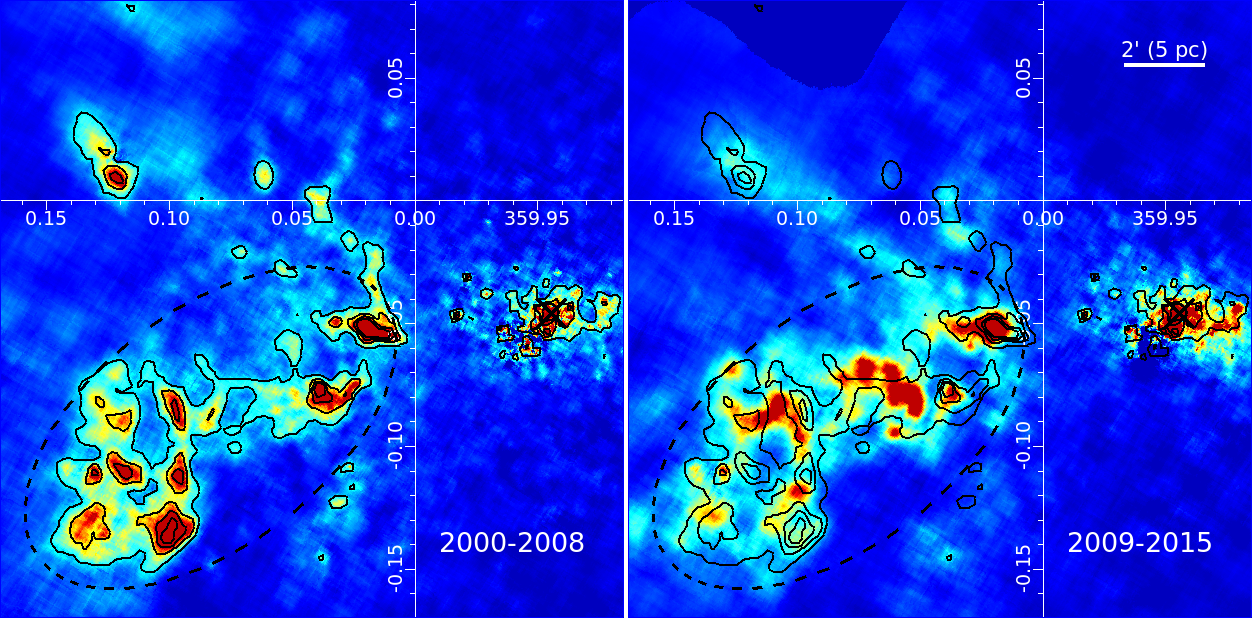}
\caption{Comparison of Chandra images of the reflected emission
  averaged over 2000-2008 (left) and 2009-2015 (right)
  observations. The maps shown are in Galactic coordinates.
  It is clear that (i) there are significant spatial
  variations of the reflected flux and (ii) there are significant
  temporal variations on time scales of few years. The maps are based
  on spectral decomposition of the data in the 5-8 keV energy band into
  reflected and hot-plasma components [see eq.~(\ref{eq:smodel})]. The position of
  the Sgr~A$^*$ is marked with a ``cross''. The dashed line shows the
  area used for the structure function calculations. Below, we refer to this region as Sgr~A complex. Note, that bright regions seen in the maps need not to be coherent structures in the velocity domain. 
\label{fig:ab}
}
\end{figure*}

In order to generate maps of reflected emission we follow the procedure described in \citet{ec16b}.
Briefly, we assume that the observed spectrum $S(E)$, in the energy band from 5 to 8 keV, can be described by a linear combination of two spectral templates: 
\begin{eqnarray}
  S(E)=A_1 R(E)+A_2 P(E),
  \label{eq:smodel}
\end{eqnarray}
where, $R(E)$ and $P(E)$ are the template spectra for the reflected
component and hot plasma emission, respectively, and $A_1$ and $A_2$
are  free parameters of the model. For the hot plasma component
$P(E)$, we use the APEC model \citep{2001ApJ...556L..91S} of an
optically thin plasma emission with a temperature fixed to 6~keV. This component is merely a convenient way to capture the contribution of faint X-ray sources to the 5-8 keV band \citep{2009Natur.458.1142R}.   For
the reflected component $R(E)$, we have selected one of the simulated
spectra of the reflected emission from \citet{ec16b}. In these
Monte-Carlo simulations a spherical homogeneous cloud is illuminated
by a parallel beam of X-ray radiation with a power law spectrum and
photon index $\sim 2$.  These models are characterized by the Thomson
optical depth of the cloud $\tau_T$ and the scattering angle (primary source -- cloud -- observer). In the
subsequent analysis, we use one of these models, 
corresponding to $\tau_T=0.5$ and the scattering angle of
90\deg. Despite the simplicity of the model given by
eq.~(\ref{eq:smodel}), the direct fitting of the spectra, extracted from
several representative regions, shows that the model captures essential
signatures of the reflected component -- 6.4 keV line and hard X-ray
continuum and separates contributions of hot plasma and reflected
components \citep[see][]{ec16b}.

A linear nature of the spectral model given by eq.~(\ref{eq:smodel})
permits to readily generate the maps of each component. Such an analysis has
been done for Chandra and XMM-Newton observations of the Galactic
Center region. As an illustration, Fig.~\ref{fig:ab} shows the maps of
the reflected component averaged over 2000-2008 (left) and 2009-2015
(right) observations with Chandra. The images have been adaptively
smoothed to reduce the photon counting noise. The comparison of these
two images shows very clearly that i) the reflected component has
a clear spatial substructure and ii) on time scales on the order of 10
years, the substructure changes very strongly. This is consistent with
previous findings
\citep[e.g.,][]{2007ApJ...656L..69M,2010ApJ...714..732P,2012A&A...545A..35C,2013A&A...558A..32C}.

The question rises about the relation between spatial and temporal
variations of the reflected component intensity. We address this
question in the next section.

\section{Structure Function in Spatial and Time Domains and l.o.s. positions of molecular clouds}
\label{sec:sf}

\begin{figure*}
\begin{minipage}{0.46\textwidth}
\includegraphics[trim= 1mm 6cm 20mm 2cm,
  width=1\textwidth,clip=t,angle=0.,scale=0.9]{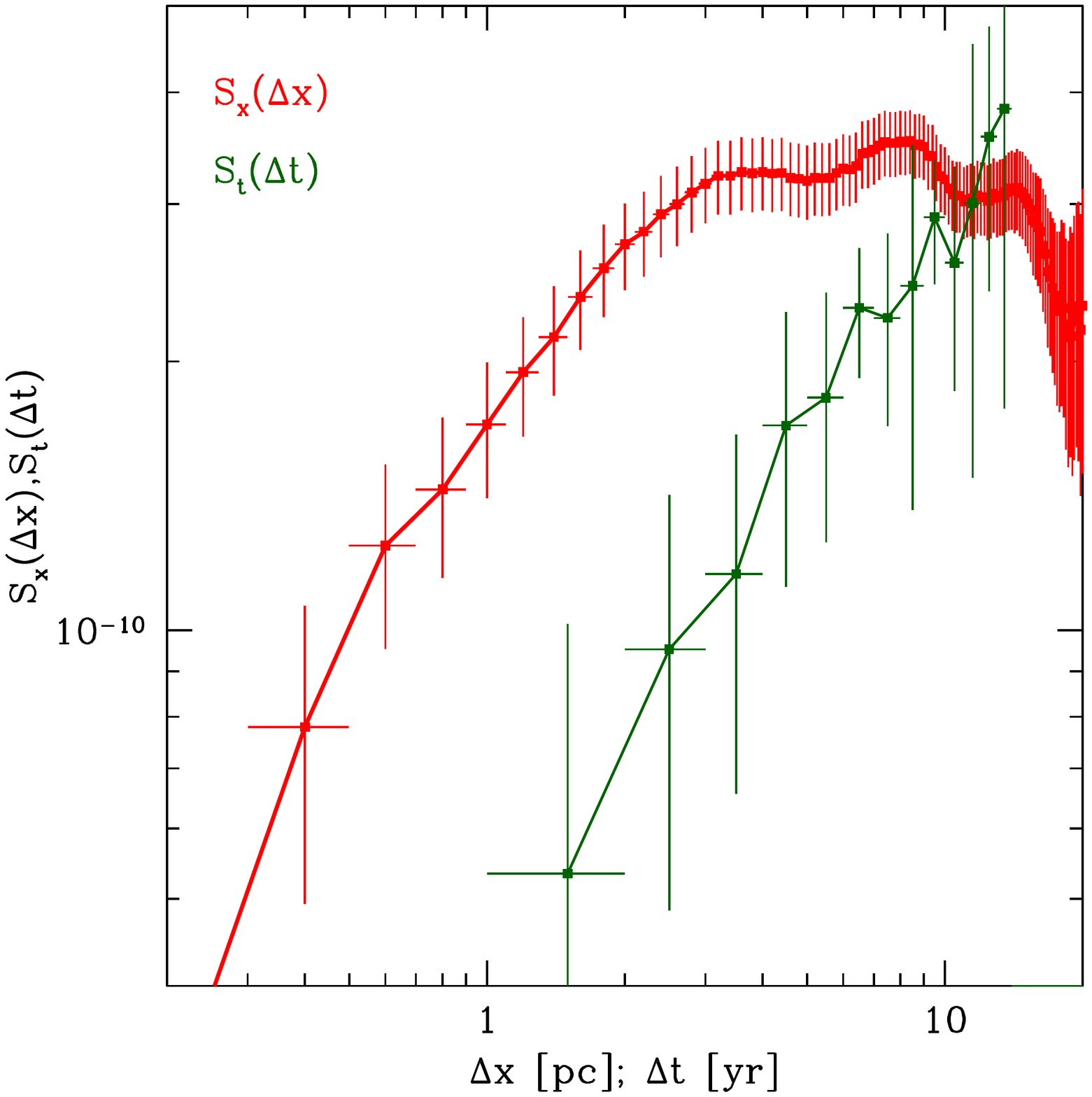}
\end{minipage}
\begin{minipage}{0.46\textwidth}
\includegraphics[trim= 1cm 6cm 20mm 2cm,width=1\textwidth,clip=t,angle=0.,scale=0.9]{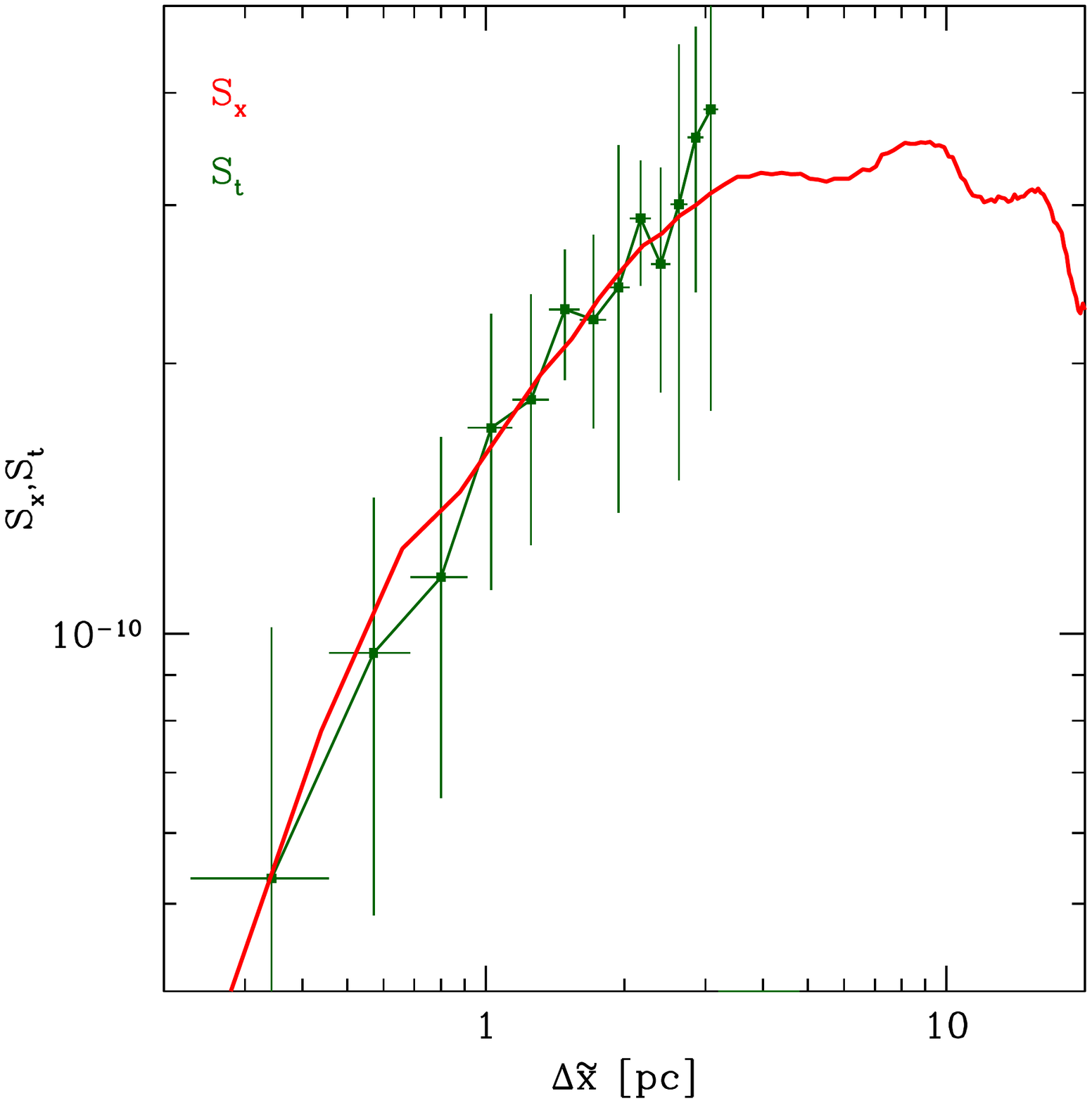}
\end{minipage}
\caption{{\bf Left:} Measured structure functions of the reflected
  component flux variations as a function of projected distance
  separation (red) and time (green). All publicly available XMM-Newton
  observations of the Sgr~A complex were used. The measurements are
  done within the region shown in Fig.~\ref{fig:ab} with the dashed
  line. The expected Poisson noise contribution has been subtracted.
  In the reflection scenario the structure functions are not
  independent, but are closely related. If the coluds are illuminated
  by a short flare of X-ray emission from Sgr~A$^*$, then one can
  simply shift these function along the horizonal axis to infer the
  age of the outburst.  {\bf Right:} Comparison of the ``shifted''
  time and space structure functions under assumption that there was a
  single and short outburst from Sgr~A$^*$ about 110 years ago. Good
  agreement suggests that the data are consistent with this
  scenario. If correct, the short flare scenario predicts that the structure
  function in the time domain should flatten at $\Delta t \gtrsim 15-20$~yr
  to match the flattening of the spatial structure function. This
  prediction can be verified with future observations covering much
  longer period of time. At present the uncertainties on the structure
  function $S_t(\Delta t)$ at $\Delta t \sim 15$~yr are far too large
  to place meaningful constraints.
\label{fig:sf}
}
\end{figure*}

In order to study spatial and temporal variations, we used all publicly
available XMM-Newton observations of the Sgr~A$^*$ region. Compared to
Chandra, XMM-Newton has larger FoV and more regular (in time) coverage of the GC
region.  Applying the procedure described in the previous section to
individual XMM-Newton observations we got a set of reflection maps
$I(\vec{x},t)$, where $\vec{x}=(x,y)$ are the projected coordinates and $t$ is the time of the observation. To
characterize variations of the flux in time and space, we calculated the
structure functions $S$, defined as
\begin{eqnarray}
S_s(\Delta \vec{x})=\left \langle \left [ I(\vec{x},t)-I(\vec{x}+\Delta \vec{x},t)\right ]^2\right \rangle,
\label{eq:sfx}
\end{eqnarray}
\begin{eqnarray}
S_t(\Delta t)=\left \langle \left [ I(\vec{x},t)-I(\vec{x},t+\Delta t)\right ]^2\right \rangle,
\label{eq:sft}
\end{eqnarray}
for the space and time domain respectively, where averaging is done over $t$ and $\vec{x}$. The resulting structure functions are shown in Fig.~\ref{fig:sf} (left panel). The Poisson noise contribution has been removed from the plotted data points. 

What can we learn from the comparison of $S_x$ and $S_t$? Let us first
examine the relation between the structure function of the gas density
distribution and the reflected flux structure functions.

Assuming that at energies above 5 keV, the gas is optically thin, i.e.,
optical depth $\tau\ll 1$, the observed reflected X-ray flux
$I(\vec{x},t)$ is
\begin{eqnarray}
I(\vec{x},t)\propto \int  \rho(\vec{x},z)\frac{f(\theta)}{r^2} L(t') dz,
\label{eq:i}
\end{eqnarray}
where $\rho(\vec{x},z)$ is the gas density, $z$ -- the coordinate
along the line of sight,  $r=\left ( R^2+z^2\right )^{1/2}$, $f(\theta)$ accounts for
the angular dependence of the reflected emission on the scattering
angle $\theta$ (for the fluorescent line alone $f(\theta)={\rm
  const}=1/4\pi$), $L(t')$ is the luminosity of the primary source and
\begin{eqnarray}
z=\frac{c}{2}(t-t')-\frac{R^2}{2c(t-t')},
\label{eq:z}
\end{eqnarray}
 \citep[see, e.g.,][]{1939AnAp....2..271C}, where $R=|\vec{x}|$ is the projected distance from Sgr~A*. 

\begin{figure}
  \begin{minipage}{0.49\textwidth}
    \includegraphics[trim= 0.5cm 3cm 0.5cm 3cm, width=1.0\textwidth,clip=t,angle=0.,scale=1.0]{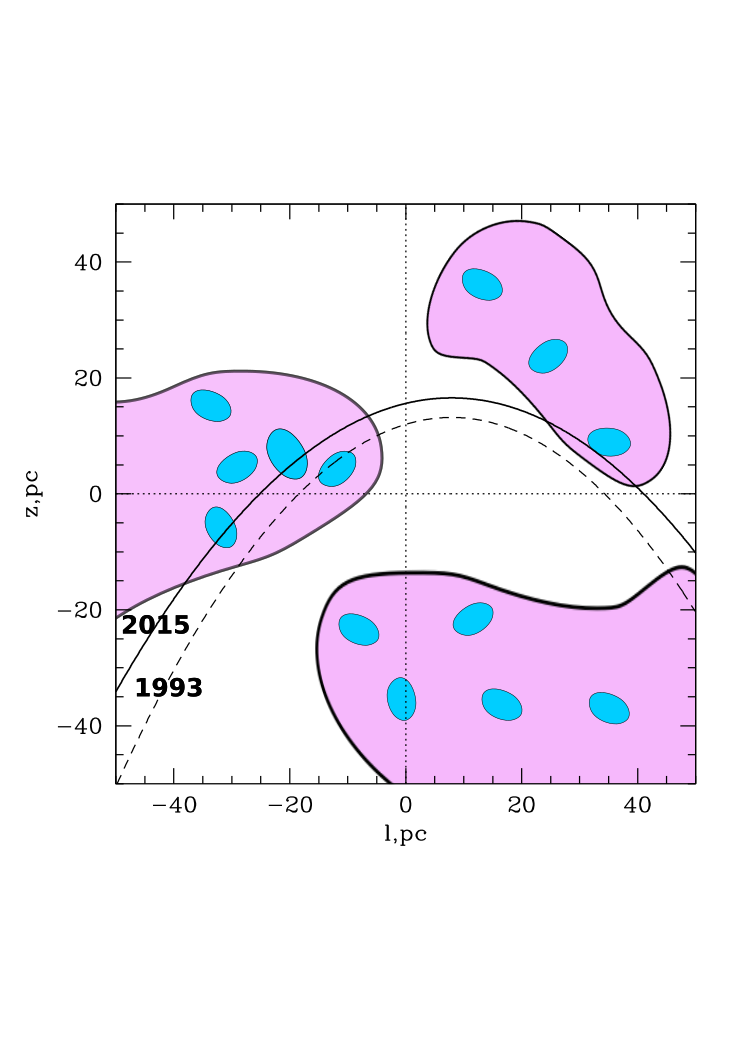}
  \end{minipage}
  \caption{Sketch of a short outburst scenario for the outburst that
    happened $\sim 110$ years ago, assuming that the primary source is
    at (0,0). This is a view on the Galactic Center region from the
    Galactic pole, with the galactic longitude $l$ along the
    horizontal axis, and the line of sight distance $z$ along the
    vertical axis. The dashed and solid lines correspond to the
    position of the ``light-front'' in 1993 and 2015,
    respectively. The distance along the line of sight between the
    illuminated regions is given by $\sim\frac{\partial z}{\partial
      t}(t_2-t_1)$, where $t_2-t_1\sim 22~{\rm yr}$ (see text). For
    regions close to primary source in projection, the value
    $\frac{\partial z}{\partial t}\sim 0.5c$. This value increases
    with projected distance from the primary source.
    \label{fig:geo}
    }
\end{figure}

The sketch of the geometry is shown in Fig.~\ref{fig:geo}.  For known
$\rho(\vec{x},z)$ and $L(t')$ the value of $I(\vec{x},t)$ can be
readily calculated. The structure function includes averaging over
many points, in space and time, and in order to predict it, it is
sufficient to know the power spectrum $P_I(k_x,k_y,k_t)$ of
$I(\vec{x},t)$, since $S(\vec{\Delta})\propto \int {P_I(k_x,k_y,k_t)
  (1-\cos 2\pi \vec{\Delta} \vec{k})d^3k}$. The power spectrum of
$I(\vec{x},t)$ can be found using eq.~(\ref{eq:i}) via power spectra
of $\rho$ and $L$.  For our purposes, it is convenient to consider
a simplified version of eq.~(\ref{eq:i}), assuming that i) we are dealing
with a spatially small patch of the image centered at
$\vec{x}_0=(x_0,y_0)$ and ii) the outburst was short, i.e., we can set
$L(t')=\Phi \delta(t'-t_0)$, where $t_0$ is the time when the outburst
occurred and $\Phi$ is the total energy emitted.  It is convenient to
choose the coordinate system for this patch in such a way that $x$ varies 
along the line connecting the primary source and the cloud, and $y$ varies in
the orthogonal direction, and $y_0=0$ chosen at the center of the patch,
i.e., $R=x_0$, $r=\sqrt{x_0^2+z_0^2}$. With these approximations one can
assume that $\frac{f(\theta)}{r^2}\approx{\rm const}$ and $\displaystyle
I(\vec{x},t)\propto \rho(\vec{x},z)$, where $z$ obeys
eq.~(\ref{eq:z}), substituting $t'=t_0$.

We now make one more important assumption that the structure function
$S_\rho$ of the density variations described by $\rho(\vec{x},z)$ is
isotropic in space on scales smaller than several pc\footnote{On large scales (larger than $\sim$10 pc) this assumption is likely violated.}. The relation between $S_\rho$ and the measured
structure functions is as follows
\begin{eqnarray}
  S_t(\Delta t)&=&AS_\rho\left(\Delta t \frac{\partial z}{\partial t}\right)\nonumber \\
 \label{eq:sfrel}
S_x(\Delta x)&=&AS_\rho\left(\Delta x \sqrt{1+\left(\frac{\partial z}{\partial x}\right)^2}\right )\\
S_y(\Delta y)&=&AS_\rho\left(\Delta y \right),\nonumber
\end{eqnarray}
 where $A$ is a constant (same for all expressions) and
 \begin{eqnarray}
\frac{\partial z}{\partial t}&=&\frac{c}{2}\left [1+\left ( \frac{x_0}{c(t-t_0)}\right)^2\right ] \nonumber \\
\frac{\partial z}{\partial x}&=&-\frac{x_0}{c(t-t_0)}
\label{eq:der} 
\end{eqnarray}

The physical interpretation of eq.~(\ref{eq:sfrel}) is clear from
Fig.~\ref{fig:geo}: with time the locus of the illuminated points at a
given $(x,y)$ is moving in $z$ direction at the velocity
$\displaystyle \frac{\partial z}{\partial t}$. Thus, our measurements
of time variations of the reflected flux are effectively probing
variations of density along $z$. The variations of flux along $x$ are
affected by the inclination of the ``parabaloid'' to the picture
plane, consequently, in projection we see clouds ``squeezed'' in this
direction (see also \S\ref{sec:dis} and Fig.~\ref{fig:xy} below). Finally, $S_y$ is a direct probe of $S_\rho$.

From eq.~(\ref{eq:sfrel}) it is clear that by comparing structure
functions in time and space domains, we can determine, e.g., the value
of $\frac{\partial z}{\partial t}$, which in turn [see
  eq.~(\ref{eq:der})] relates $x_0$ and the time elapsed after the
outburst. Since our spatial structure function has been calculated
without making any distinctions between $x$ and $y$ direction, we
adopt a simplified expression for $S_s(\Delta)$:
\begin{eqnarray}
S_s(\Delta)&=&AS_\rho\left(\Delta \left[\frac{1}{2}\sqrt{1+\left(\frac{\partial z}{\partial x}\right)^2}+\frac{1}{2}\right]\right ),
\label{eq:ss}
 \end{eqnarray}
 intermediate between $S_x$ and $S_y$, given by eq.~(\ref{eq:sfrel}). 
 
 From eq.~(\ref{eq:sfrel}) it follows that spatial and time structure
 functions should coincide if they are plotted with respect to the
 effective $\tilde{\Delta}_x$, i.e.
  \begin{eqnarray}
  \tilde{\Delta}_x&=&\Delta_x \frac{1}{2}\left[\sqrt{1+\left ( \frac{x_0}{c(t-t_0)}\right)^2}+1\right]\\
    \tilde{\Delta}_x&=&\Delta_t \frac{c}{2}\left [1+\left ( \frac{x_0}{c(t-t_0)}\right)^2\right ]
  \end{eqnarray}
for the space and time domain respectively.  We now can search for the
value of $\eta=\frac{x_0}{c(t-t_0)}$ that provides the best match
between two structure functions. This exercise is illustrated the in right
panel of Fig.~\ref{fig:sf}. The best match is found for $\eta=0.7$,
adopting $x_0\approx 10' \approx 27.3$~pc, implying that the outburst
happened $t-t_0\approx 110$~yr ago. Another immediate conclusion is
that bright clouds are located $z\sim 10$~pc further away than
Sgr~A$^*$.

Within $\sim8$~pc from Sgr~A$^*$ (in projection) there are two
prominent molecular structures, the so-called 50~${\rm km~s^{-1}}$ and
20~${\rm km~s^{-1}}$ clouds.  There is no evidence for bright
reflected emission from these clouds in the Chandra and XMM-Newton data.
\citet{2000ApJ...533..245C} suggested that these clouds are located
within $z\pm 10$~pc, but on the opposite sides (along the
line-of-sight) from Sgr~A$^*$. Since for a closer vicinity of Sgr~A$^*$
the locus of illuminated gas is located at $z\sim 0.5c\times 110~{\rm
  yr}\sim 17$~pc, the flare scenario predicts that by now both clouds
are not illuminated -- consistent with X-ray data. The 50~${\rm
  km~s^{-1}}$ could have been bright $\sim 35~$yr ago (if located at
$z=10~$pc) or even earlier.

\begin{figure*}
\begin{minipage}{0.49\textwidth}
\includegraphics[trim= 1mm 5cm 20mm 2cm, width=1\textwidth,clip=t,angle=0.,scale=0.9]{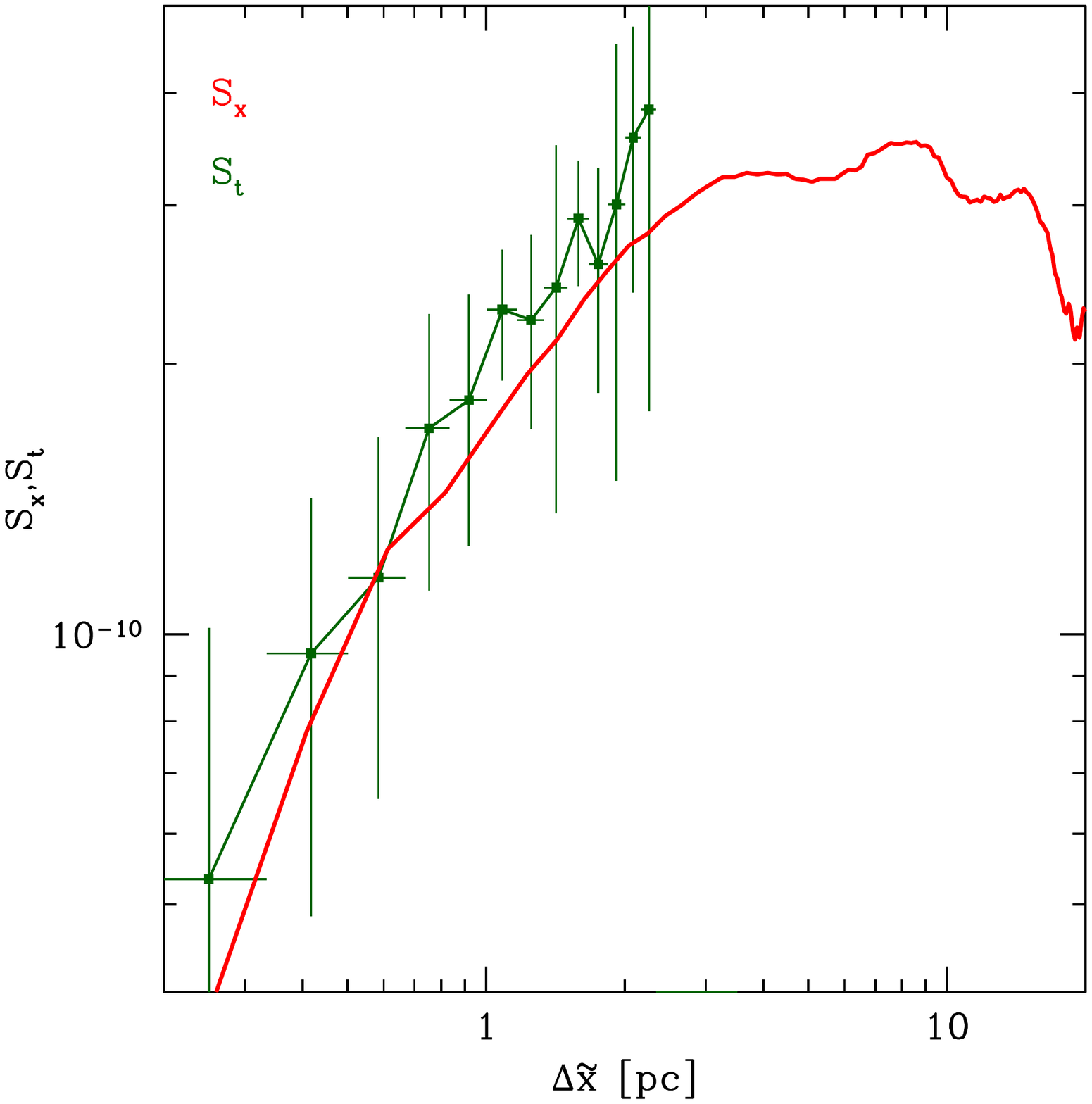}
\end{minipage}
\begin{minipage}{0.49\textwidth}
\includegraphics[trim= 1mm 5cm 20mm 2cm,width=1\textwidth,clip=t,angle=0.,scale=0.9]{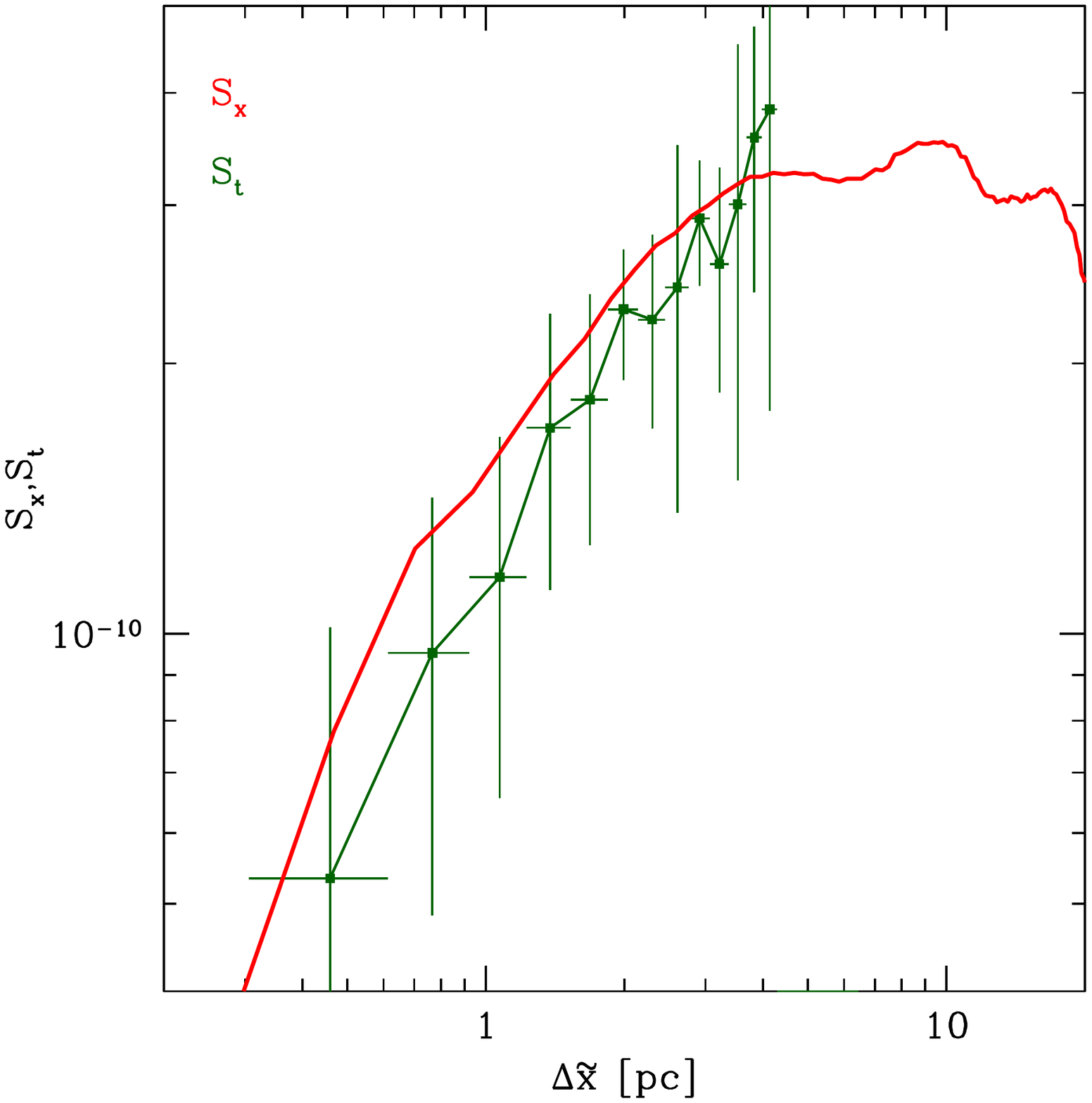}
\end{minipage}
\caption{Same as in Fig.~\ref{fig:sf} for $\eta=0.3$ (left) and
  $\eta=1$ (right). In terms of elapsed time since the outburst the
  figures correspond to $\sim 250$ and $\sim 75$~yr
  respectively. Clearly the agreement between the structure functions is
  much poorer than for $\eta=0.7$.
\label{fig:ex}
}
\end{figure*}

Experiments with other values of $\eta$ have shown that the values of
$\eta$ between 0.65 and 0.8 provide similarly good
match. The corresponding range of the $t-t_0$ is $\sim$120-95~yr. On the
other hand, for a broader range of values, e.g., $\eta=0.3$ or 1, there is a
clear mismatch between the manipulated structure functions (see
Fig.~\ref{fig:ex}).

\begin{figure*}
\begin{minipage}{0.45\textwidth}
\includegraphics[trim= 1mm 2cm 0mm 2cm, width=1\textwidth,clip=t,angle=0.,scale=0.9]{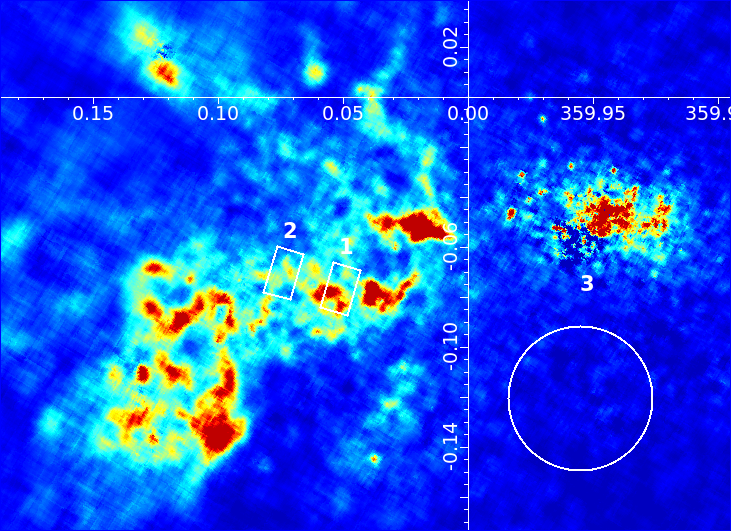}
\end{minipage}
\begin{minipage}{0.49\textwidth}
\includegraphics[trim= 1cm 4cm 20mm 2cm,width=1\textwidth,clip=t,angle=0.,scale=0.9]{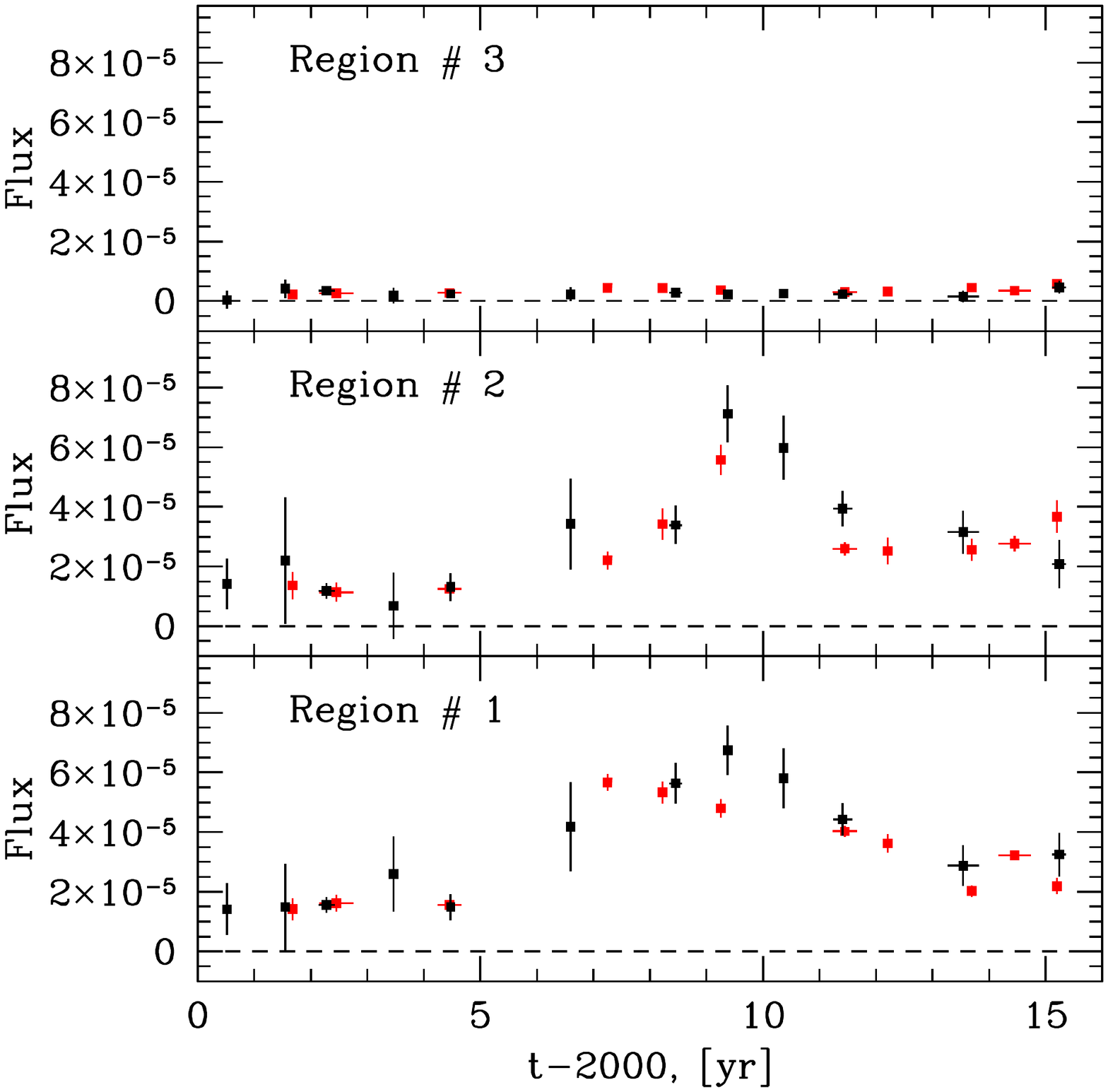}
\end{minipage}
\caption{Light curves of reflected emission for selected regions. {\bf
    Left:} Three regions used for the analysis, superposed onto
  Chandra map. Regions 1 and 2 are located within the molecular
  complex bright in reflected emission. Region 3 is a test region
  where reflected emission is faint. {\bf Right:} Light curves for
  regions 1,2 and 3 (from bottom to top). Red and black points
  correspond to XMM-Newton and Chandra observations respectively.  The
  duration of the peaks sets an upper limit on the duration of the
  outburst. Even for an extremely short outburst the finite spatial
  extent of each cloud along the line of sight broadens the peak. For
  instance, for a cloud with size of 1~pc, the minimum duration of the
  observed outburst, given $\frac{\partial z}{\partial t}\sim 0.75 c$
  is 4.3~yr.
\label{fig:xvar}
}
\end{figure*}

The above calculations were done assuming that the duration of the
outburst $t_b$ is infinitely small. One gets constraints on $t_b$
directly from the structure functions, since at small separations in
time the function $S_t(\tilde{\Delta}_x)$ should drop compared to
$S_s(\tilde{\Delta}_x)$. However the difference between
$S_t(\tilde{\Delta}_x)$ and $S_s(\tilde{\Delta}_x)$ will only be prominent
if the duration of the outburst is significantly longer than the
light propagation time of a typical cloud along the line of sight,
i.e., $t_b\gtrsim l/\frac{\partial z}{\partial t}$, where $l$ is the
characteristic size of the cloud. In addition, the finite angular
resolution of the XMM-Newton is expected to introduce by itself a drop
of the $S_s(\tilde{\Delta}_x)$ at small separations. If we assume an
effective angular resolution of $\sim10''$, it translates into
$\sim0.4$~pc in space, or $\sim1.7$~yr in time (assuming
$\frac{\partial z}{\partial t}=0.75c$).  From Fig.~\ref{fig:sf} there
is no evidence for a strong decline of $S_s$ w.r.t. $S_t$. This
suggests that the duration was shorter than a few years \citep[see also][]{2013A&A...558A..32C}. One can
corroborate this conclusion by examining light curves of a few variable
bright spots in the reflected emission maps (see Fig.~\ref{fig:xvar}). The light
curves have been extracted from three regions. Regions 1 and 2 are
located within the part of the Sgr~A molecular complex that is bright in the reflected emission,
while region 3 is a test region where reflected emission is
faint. Corresponding light curves are shown in the right panel of
Fig.~\ref{fig:xvar}.  Red and black points correspond to XMM-Newton
and Chandra observations, respectively. The flux is clearly variable
with a peak in 2009-2010. The duration of the peaks sets an upper
limit on the duration of the outburst. Even for an extremely short
outburst the finite spatial extent of each cloud along the line of
sight broadens the peak in the light-curves. For instance, for a cloud with the size of 1~pc,
the minimum duration of the observed outburst, given $\frac{\partial
  z}{\partial t}\sim 0.75 c$ is 4.3~yr. Based on this analysis, we have
concluded that the assumption of a short outburst [shorter than a few
years; see also \citet{2013A&A...558A..32C}] is reasonable. 
  
\section{Discussion} 
\label{sec:dis}
The mean surface brightness of the reflected component in the analysed
region (see Fig.~\ref{fig:ab}) is $\sim 2~10^{-5}~{\rm
  phot~cm^{-2}~s^{-1}~arcmin^2}$ in the 5-8 keV band.  From this value (assuming fiducial
parameters derived in the previous section and isotropic reflected
emission) we can immediately estimate the total fluence $\Phi=L\times
t_b$ of the primary source during the outburst
\begin{eqnarray}
  \Phi_{1-100~{\rm keV}}\sim 10^{48}~\rho_3^{-1}{\rm erg},
  \label{eq:flu}
\end{eqnarray}
where $\rho_3$ is the mean gas density within the illuminated region
in units of $10^3~{\rm cm^{-3}}$. The fluence in the 1-100~keV band
was recalculated from the measured 5-8 keV flux assuming a photon
index of 1.8, as estimated from INTEGRAL observations of the Sgr B2
cloud \citep[][]{2004A&A...425L..49R}. Corresponding conversion factor
is $\sim 9.3$. For a photon index of 2
\citep[see][]{2010ApJ...719..143T,2015ApJ...814...94M} this factor is
$\sim$9.8. A fiducial value $\rho=10^3~{\rm cm^{-3}}$ used in
eq.~(\ref{eq:flu}) was motivated by the well studied Sgr~B2 complex,
where $10^3~{\rm cm^{-3}}$ gas dominates the total mass budget on
scales of tens of pc \citep[][]{2016A&A...588A.143S}. The size of the
region studied here (Fig.~\ref{fig:ab}) is $\sim 10$~pc. Thus it seems
reasonable to use the same fiducial value of density as in Sgr~B2.
The surface brightness in the studied region varies by a factor 10
(lower limit if small-scale substructure is present). Given our
fiducial value of mean density, the brightest regions would correspond
to local densities $\sim10^4~{\rm cm^{-3}}$.

The luminosity,
corresponding to the estimated fluence is
\begin{eqnarray}
  L_{1-100~{\rm keV}}\sim 5~10^{40}~\rho_3^{-1} ~\left ( \frac{t_b}{1{\rm ~ yr}}\right )^{-1}{\rm erg~s^{-1}}.
  \label{eq:lum}
\end{eqnarray}
Note that in the {\it Nustar} study of the same region,
\citet{2015ApJ...814...94M} suggested a much weaker limit on the
Sgr~A$^*$ luminosity $\gtrsim 10^{38}~{\rm erg~s^{-1}}$ in the 2-20
keV band. Apart from the factor of $\sim 2$ difference due to narrower
energy band, the remaining difference is due to (i) the way the
density/optical depth is defined and (ii) the assumed outburst
duration. In \citet{2015ApJ...814...94M} the column density $N_H\sim
10^{23}~{\rm cm^{-2}}$ of the reflected gas was estimated from the
spectral modelling. For our fiducial value of $\frac{\partial
  z}{\partial t}\sim 0.75c$ and assumed duration of $t_b\sim$1~yr, the required
density is $N_H/(\frac{\partial z}{\partial t}t_b)\sim 10^{5}~{\rm
  cm^{-3}}$. The column density measurements from the spectra rely
on the spectral distortions introduced by photoelectric absorption. However, if
only a part of the cloud is illuminated, as it is the case when the flare is
very short, the spectrally-derived column density corresponds to
the characteristic column density of the entire cloud, rather than the
column density of the illuminated gas. We therefore conclude that in
the case of a short flare the column density (or mean density) that
enters the conversion from the surface brightness to the luminosity of
the primary source [see eq.~(\ref{eq:lum})] does not necessarily
coincide with the spectrally-determined column density.  We therefore
keep our scaling (via $\rho_3$) in all subsequent estimates, but
acknowledge that this value is uncertain by $\sim$ one order of magnitude.

Based on Chandra observations of molecular clouds (similar region to
the one studied here), \citet{2013A&A...558A..32C} suggested that apart
from a short ($\sim$2~yr long) flare that dominates the variability of
some clouds, additional flares are responsible for the illumination of
other clouds. This conclusion is primarily based on the comparison of
the maximal fluxes from a set of clouds and the peak antenna
temperatures in the CS and N$_2$H$^{+}$ lines for the same
clouds. \citet{2013A&A...558A..32C} found that the ratio of the X-ray
and line fluxes differ among five clouds by a factor of $\sim$6. We
believe that the uncertainties in association of X-ray peaks with the peaks from the
position velocity cubes and conversion of molecular line fluxes to gas
densities/column densities are large enough to allow for a single outburst
scenario.  We therefore keep the assumption of a single
and short flare for the rest of the paper since it leads to the
simplest picture. We note in passing that the structure functions can
be explicitly expressed through the convolution of the gas density
distribution power spectrum and the power spectrum of the flare. A
short flare model, used here, has the advantage that most of the
calculations are reduced to simple algebraic manipulations.

%-----------------------
The minimal accreted mass, assuming high radiative efficiency at
  level of 10\%, is $\delta M\sim \phi/(0.1c^2)\sim
  10^{28}~\rho_3^{-1}~{\rm g} \sim 10^{-5}~\rho_3^{-1}~M_\odot \sim
  10^{-2}~\rho_3^{-1}~M_{J}\approx 0.3~M_{E}$, where $
  M_{J}\approx10^{-3}~M_\odot$ and $ M_{E}\approx
  3\times10^{-6}~M_\odot$ are Jupiter and Earth masses,
  respectively. Thus, this is a rather small value that can be easily
  provided by a tidal disruption (TDE) of a planet, or capture of the gas cloud similar to the one found by \citet{2012Natur.481...51G}. A scenario with a
  Jupiter-mass planet has been already considered in application to
  the past Sgr~A$^*$ activity by \cite[][]{2012MNRAS.421.1315Z}, but
  the predicted duration of a flare in this case is likely $ \gtrsim
  10 $ yrs, while the actual frequency of such events is highly
  uncertain. 

In order to produce a shorter flare one needs to disrupt a more
massive body (having higher escape velocity), i.e. a super-Jupiter
planet \citep[see, e.g.,][]{2002ApJ...576..753L,2012MNRAS.421.1315Z,2013A&A...552A..75N}
or a star. The rate of ``canonical'' stellar TDEs is estimated at
level $ \sim \rm{few} \times 10^{-5}~yr^{-1} $ per galaxy \citep[][]{2014ApJ...792...53V,2014MNRAS.444.1041K}, while the
peak luminosity of an associated flare can be as high as the Eddington
limit for the SMBH \citep{1988Natur.333..523R,1989ApJ...346L..13E,1999ApJ...514..180U}.
However, the bulk of this radiation is likely to be in extreme UV and soft X-ray energy bands\citep[][]{2009ApJ...698.1367G,2015JHEAp...7..148K}, while hard X-ray emission is expected to originate from either a relativistic jet \citep{2011Sci...333..203B,2011Natur.476..421B,2012ApJ...753...77C} or a corona of hot electrons. In the former case, the
hard X-ray emission should be beamed in the jet's direction. Therefore, if
viewed off-axis, the source can appear significantly dimmer. In the
latter case, hard X-ray emission would also be rather dim due to the small
optical thickness of the corona. Both of these opportunities
still imply that the hard X-ray emission constitutes only a
minor fraction of the event's total energy output, and therefore some
manifestations of its major contributor (i.e. soft X-ray emission,
beamed X-ray emission or kinetic power of the jets) should be present
(e.g. signatures of severe X-ray heating of the gas or jet-driven
shocks). Still, some recent studies predict that a much smaller total
radiative output might be more typical for TDEs
\citep[][]{2015ApJ...806..164P}, as a result of inefficient
circularization of bound debris of the disrupted star.  All these
scenarios remain speculative, given the uncertain, but in general low
frequency of TDEs \citep[see however][for discussion of a
phenomenological model, extrapolating the Sgr~A$^*$ flaring activity to
rare, but very bright events]{2012ApJS..203...18W}. 

%c------------------- refs

%c-------------------------------------

If Sgr~B2 is exposed to the same outburst, than it is located at
$z\sim -130$~pc (closer to us than Sgr~A$^*$), and the velocity of the flare propagation along the line of sight is $\frac{\partial
  z}{\partial t}\sim 5c$ (for our fiducial parameters). The position estimate
coincides with the value derived by \citet{2009ApJ...705.1548R} based
on the assumption that Sgr~B2 is on the low eccentricity orbit around the 
Galactic Center. While the equality of two estimates is of course a
coincidence, given the assumption and uncertainties, it is tempting to
further examine this scenario. Firstly, one can consider this
result as a confirmation that Sgr~B2 is on the low eccentricity
orbit. Secondly, we can directly compare the gas densities in the Sgr~B2 cloud (and in any other cloud) relative to the Sgr~A complex. The surface brightness of the  reflected emission power by a short outburst is
\begin{eqnarray}
  I\propto \rho\frac{\partial z}{\partial t}r^{-2}\propto \rho \left [1+\left ( \frac{x}{c(t-t_0)}\right)^2\right ] \left[x^2+z^2 \right]^{-1},
  \label{eq:ix}
\end{eqnarray}
where $z$ is related to $x$ and $t-t_0$ according to eq.~(\ref{eq:z}).
Compared to the Sgr~A complex, the decrease of the
surface brightness due to $r^{-2}$ term is partly compensated by the
larger $\frac{\partial z}{\partial t}$ term ($\sim5c$ for Sgr~B2
compared to $\sim0.75c$ for the Sgr~A complex). During recent years, (e.g., 2012 observations
of XMM-Newton), the mean surface brightness of the Sgr~B2 region in the
reflected emission was $\sim5-6$ times lower than the mean surface
brightness in the Sgr~A complex (Fig.~\ref{fig:ab}), which is consistent with
the predictions. During earlier observations, e.g., before 2002, the
two molecular complexes were almost equally bright, but this can be
readily explained in the single-flare scenario if during these
observations a denser part of the Sgr~B2 complex was illuminated. Yet
another constraint could come from the duration of the period when
Sgr~B2 was bright -- at least for 24 years (from 1993 to 2015). Given that
our fiducial value of $\frac{\partial z}{\partial t}\sim 5c$, this
requires a size of $5c\times 24~{\rm yr}\approx 37$~pc along the
line of sight. This is consistent with the size of the Sgr~B2
cloud which according to \citet{2016A&A...588A.143S} is $\sim 45$~pc.

\begin{figure*}
\includegraphics[trim= 0mm 0cm 0mm 0cm, width=1\textwidth,clip=t,angle=0.,scale=0.9]{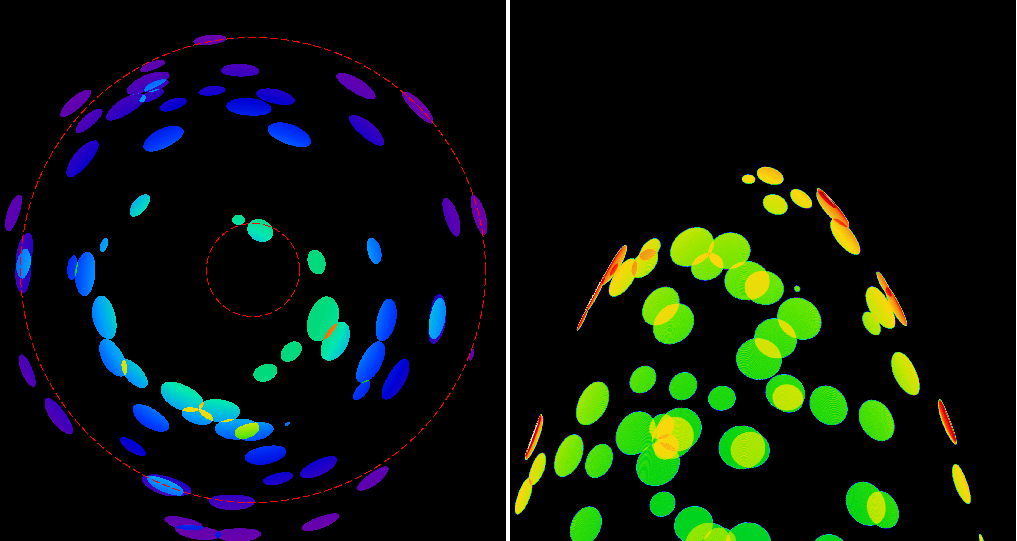}
\caption{Apparent distortions of randomly distributed spherical clouds
  illuminated by a very short flare. In this illustration each cloud has a radius of   6~pc and a uniform density. The ``paraboloid'', corresponding to the
  short flare, illuminates a thin slice in every cloud that it
  intersects. For a cloud at sufficiently large projected distance
  from the primary source, such a slice is inclined to the picture plane and in projection it looks flattened in the radial direction as shown in the
  left panel \citep[see also Fig.3 in][]{1998MNRAS.297.1279S}. The colors
  reflect the expected surface brightness of the clouds [see
    eq.~(\ref{eq:ix})]. Two circles have the radii of 20 and 100 pc,
  approximately corresponding to Sgr~A and Sgr~B2 complexes,
  respectively. While the clouds look squeezed in radial direction, in
  tangential direction the clouds are not distorted. The analysis
  of such distortions can be used to infer the position of the primary
  source and the angle of the paraboloid with respect to the line of
  sight (or, equivalently, the time of the flare).  The left panel shows  a``side view'' of the clouds that intersect with the paraboloid.  Our fiducial value
  $t_b=110$~yr is used.
\label{fig:xy}
}
\end{figure*}

The best verification of the short flare scenario could come from the
comparison of light curves from several compact cores in two molecular
complexes at different projected distances from Sgr~A$^*$ (e.g., Sgr~B2 vs Sgr~A complexes) - the shapes of the light curves
should be similar once a correction on the factor $\frac{\partial
  z}{\partial t}$ is made. Another possibility is to use the
difference between the spatial structure functions in the radial and
tangential directions [see eq.~(\ref{eq:sfrel})]. We illustrate this
type of diagnostics in Fig.~\ref{fig:xy}. The figure shows that when
spherical clouds are illuminated by a short flare, they appear
flattened in radial direction (reminiscent of weak lensing image
distortions). The magnitude of this effect increases with the projected
distance. Thus, such distortions (after averaging over many clouds)
could be used to infer the position of the primary source
[perpendicular to the squeezed direction] and the inclination of the
paraboloid with respect to the picture plane [from the magnitude of
  distortions]. As discussed in the previous section, the inclination is
a proxy for the time of the outburst. One can also use the same approach
to determine the duration of the flare - when the entire volume of a small cloud (smaller,
than $\frac{\partial z}{\partial t}t_b$) is illuminated by the flare,
the cloud should not appear as ``squeezed'' in radial
direction, while larger clouds should.

Recently, several plausible models of the 3D distribution of 
molecular gas in the GC region have been suggested
\citep[e.g.,][]{2011ApJ...735L..33M,2015MNRAS.447.1059K,2016MNRAS.457.2675H}. In
these models a significant fraction of molecular gas is attributed to
close or open orbits in the GC potential. One can use X-ray data to
verify and calibrate these models. We simulate the expected reflected
signal based on these models in another paper \citep{ec16b}. Here, we
only note that in the most plausible model of
\citet{2015MNRAS.447.1059K} the orbit passes some 50~pc behind
Sgr~A$^*$. This implies that the reflected emission that we see in X-rays
today is not associated with the clouds on this orbit. If an additional
significant mass is indeed present along this part of the orbit, then
the reflected signal from the same outburst should be seen hundreds of
years later.

In this study, we have made a number of simplifying assumptions. Some
of these simplifications can be avoided, but some are more
fundamental. Among the latter is the assumption that the power
spectrum of gas density fluctuations is isotropic. In reality, clouds
might be preferentially stretched or squeezed in radial or
tangential directions. A related problem is the finite number of clouds
and/or strong correlation between the distribution of clouds in
3D. Most of the other simplifications can be avoided by doing more
elaborate modelling based on eq.~(\ref{eq:i}) that we defer to future
work. A great deal of uncertainties can be removed by doing further
observations. For instance, sufficiently long and regularly spaced in
time (a few months) Chandra and XMM-Newton observations (preferably by
the same telescope/detector configuration) would allow to measure the
structure functions on small scales to better constrain the duration
of the outburst and spatial structure function of illuminated
clouds. Particularly powerful diagnostics would be possible with the
detection of localized and transient spikes in the surface brightness,
corresponding to small and dense cores of molecular clouds. Even more
interesting would to associate localized peaks in X-rays with
maser sources, for which parallax measurements allow independent
distance estimates. 

Another extremely interesting possibility, is to use future
polarimetric observations \citep[see,
  e.g.][]{2002MNRAS.330..817C,2015A&A...576A..19M,2016A&A...589A..88M}
that can provide a direct measurement of the primary source position in the
picture plane (from the polarization angle) and 3D position of the
cloud (from the degree of polarization). For our fiducial parameters
the scattering angle is $\sim 60$\deg for the bright part of the Sgr~A
complex, and the expected degree of polarization is close to 60\%. The
observed reflected flux from the entire ellipse (see
Fig.~\ref{fig:ab}) is $\sim 10^{-10}~{\rm erg~cm^{-2}~s^{-1}}$ in the
5-8 keV band. Even taking into account that i) fluorescent line
emission is not polarized and ii) thermal plasma emission typically
contributes about half of the observed flux, it should be possible to
detect polarization using future imaging polarimeters, e.g. XIPE
\citep{2013ExA....36..523S} or IXPE \citep{2013SPIE.8859E..08W}. The
degree of polarization would be a solid proxy for the scattering
angle. The same degree of polarization is expected for scattering by
an angle $\theta$ and $\pi-\theta$. In reality, scattering smaller
than 90\deg implies that the outburst happened less than $x0/c\sim
77$~yr ago. Even smaller angles can be effectively excluded, since it
is unlikely that a powerful outburst from Sgr~A$^*$ was missed by
observers. Therefore, this ambiguity can be resolved.

\begin{figure*}
\begin{minipage}{0.45\textwidth}
\includegraphics[trim= 0mm 0mm 5cm 0cm, width=1\textwidth,clip=t,angle=0.,scale=0.9]{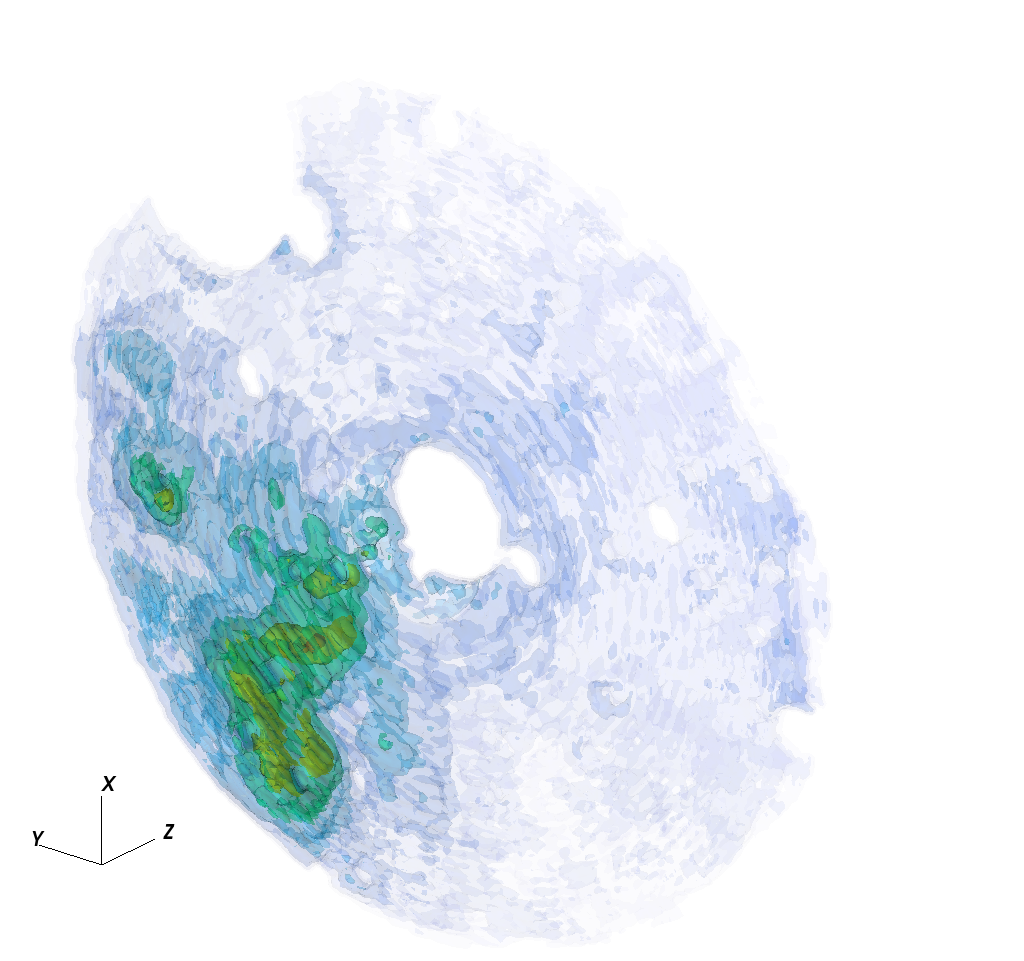}
\end{minipage}
\begin{minipage}{0.45\textwidth}
\includegraphics[trim= 0mm 0mm 5cm 0cm, width=1\textwidth,clip=t,angle=0.,scale=0.9]{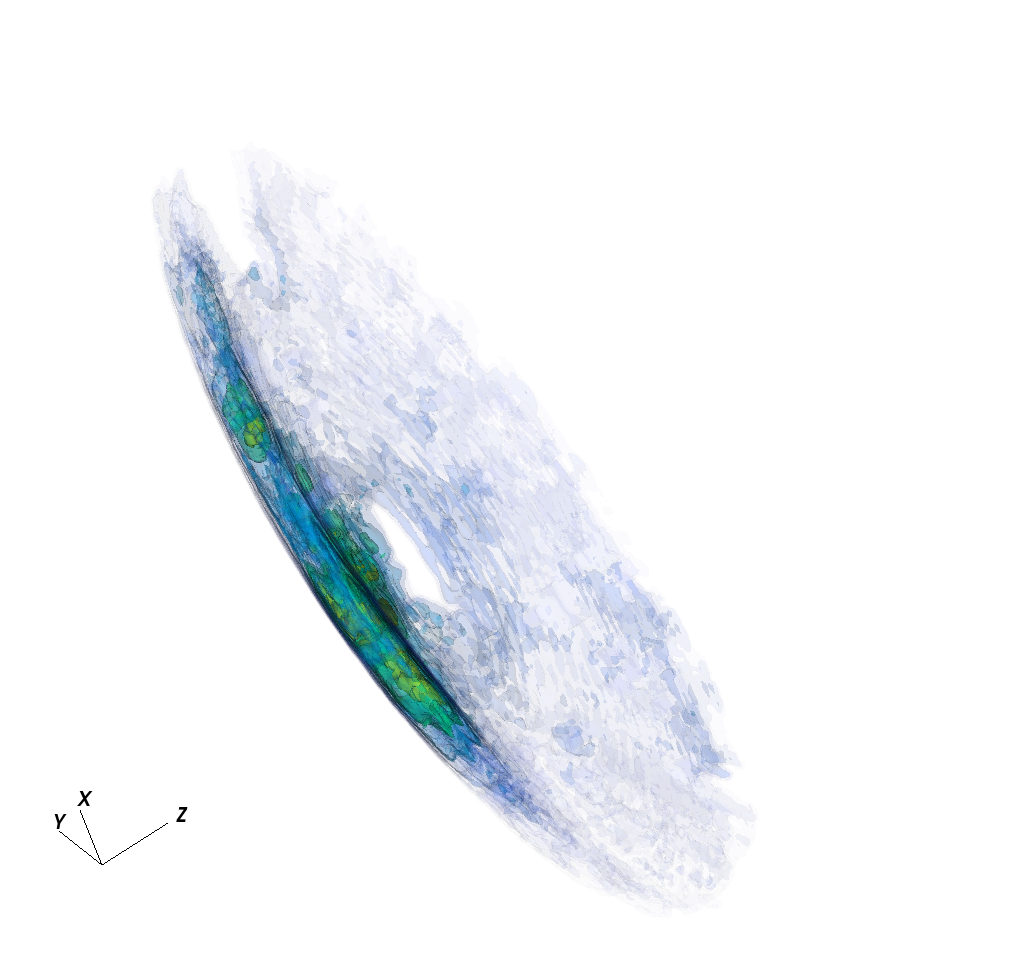}
\end{minipage}
\caption{3D distribution of gas near the Galactic Center based on our
  fiducial model of a short flare 110 years ago. The locus of the
  illuminated points at any given time is a paraboloid [see
  eq.~(\ref{eq:z})]. The surface brightness at location $(l,b)$
  observed by XMM-Newton at time $t$ is recalculated to the gas
  density (scaled by a fiducial value of total fluence) and placed to
  a cell with coordinates$(l,b,z)$. The radius of the studied area is
  $\sim 35$~pc. The “thickness” of the paraboloid along the line of
  sight covered during $\sim 15$~yr of XMM-Newton observations is
  $\sim 0.75\times c \times 15~{\rm yr}\approx 3.4$~pc.  Left and right panels show
  different orientations of the same data cube. The ``holes'' correspond to excised regions contaminated by bright compact sources.
  \label{fig:3d}}
\end{figure*}

The scenario described above opens an exciting perspective for accurate
mapping of the molecular gas density in the GC region, once the fluence of
the flare is reliably calibrated. Indeed, unlike any other method, the
conversion factor from the surface brightness of reflected emission to
the gas density is uncertain by a factor $\sim$2 at most, due to the
uncertainty in the iron abundance. All other gas properties like
temperature, ionization state (as long as iron is not strongly
ionized), molecular, atomic or dust grain phases do not make any significant
difference to the conversion factor. An ambitious program of regular
observations for {\bf a few hundred!} years would effectively provide a full
3D picture of the molecular gas distribution over the entire 100~pc region. A ``preview'' ($\sim$3.4~pc thick) of a 3D gas distribution based on XMM-Newton observations is shown in Fig.~\ref{fig:3d}. 
Moreover, once X-ray calorimeter data become available with future
observatories (like Athena, X-Ray Surveyor), this 3D mapping can be supplemented with gas velocity
measurements unambiguously linking position-velocity data from
molecular line observations with the 3D positions from X-ray data.

\section{Conclusions}
X-ray observations of the Galactic Center molecular clouds can serve
as a powerful diagnostic tool for the 3D geometry of molecular gas and
the past history of the Sgr~A$^*$ activity. We used the comparison of
spatial and temporal variations of the reflected emission flux to
conclude that $\sim$110~yr ago our supermassive black hole had a
luminosity of the order of $10^{41}~{\rm erg~s^{-1}}$ over a period no
longer than a few years. This conclusion implies that a molecular
complex, currently exposed to Sgr~A$^*$ radiation at $\sim 20$~pc from
the source (in projection), is located $\sim 10$~pc further away than
Sgr~A$^*$ along the line of sight. We argue that these estimates can
be further improved by future imaging and polarimetric
observations. If there were only one or a few outbursts of Sgr~A$^*$,
then it would eventually be possible to make a full 3D distribution of
the molecular gas in the GC region.

\FloatBarrier

\section{Acknowledgements}
The results reported in this article are based in part on data
obtained from the Chandra X-ray Observatory (NASA) Data Archive and
from the Data Archive of XMM-Newton, an ESA science mission with
instruments and contributions directly funded by ESA Member States and
NASA. We acknowledge partial support by grant No. 14-22-00271 from the
Russian Scientific Foundation.

\label{lastpage}
\end{document}